\documentclass{aa}
\usepackage{graphicx}
\begin{document}
\title{On the calibration of the COBE/IRAS dust emission reddening maps\thanks{ Based on observations made at 
Complejo Astron\'omico El Leoncito, which is operated under agreement between the Consejo Nacional de 
Investigaciones Cient\'{\i}ficas y T\'ecnicas de la Rep\'ublica Argentina and the National Universities of 
La Pata, C\'ordoba and San Juan, Argentina.}
}

\author{C.M. Dutra\inst{1,2}, A.V. Ahumada\inst{3},  J.J. Clari\'a\inst{3}, E. Bica\inst{4}, \and B. Barbuy\inst{1}}
 
\offprints{C.M. Dutra -- dutra@astro.iag.usp.br}

\institute{Universidade de S\~ao Paulo, Instituto de Astronomia, Geof\'{\i}sica e Ci\^encias Atmosf\'ericas, 
Rua do Mat\~ao 1226, Cid. Universit\'aria, 05508-900, S\~ao Paulo SP, Brazil\\
     \mail{}
     \and
Universidade Estadual do Rio Grande do Sul, Rua Bompland 512, S\~ao Borja
 97670-000, RS, Brazil\\
 \and
Observatorio Astron\'omico de C\'ordoba, Laprida 854, 5000, C\'ordoba, Argentina\\
     \mail{}
\and
Universidade Federal do Rio Grande do Sul, Instituto de Fisica, CP\,15051, Porto Alegre 91501--970, RS, Brazil\\
    \mail{}}

\date{Received ; accepted }

\abstract{ 
In this work we study the spectral properties (3600 - 6800 \AA)
of the nuclear region of early-type galaxies at low ($|b|<25^{\circ}$),
intermediate (including surroundings of the Magellanic Clouds) and high
(South Polar Cap) Galactic latitudes. We 
determine the $E(B-V)$ reddening values of the galaxies by matching their continuum 
distribution with respect to 
those of reddening-free spectral galaxy templates with similar stellar 
populations.
We also compare the spectroscopic reddening value of each galaxy 
with that derived from 
100 $\mu m$ dust emission ($E(B-V)_{FIR}$) in its line of sight, and we 
find that there is  agreement up to $E(B-V)=0.25$. Beyond this limit $E(B-V)_{FIR}$ values are higher. 
Taking into account the data up to $E(B-V) \approx 0.7$, we  derive a 
calibration factor of 0.016  between the spectroscopic $E(B-V)$ values 
and Schlegel et al.'s (1998) opacities. By combining this
 result with  an $A_K$ extinction map built within ten degrees of the Galactic centre
using Bulge giants as probes (Dutra et al. 2003), we extended the calibration of dust 
emission reddening maps to low Galactic 
latitudes  down to $|b|=4^{\circ}$ and $E(B-V)= 1.6$ ($A_V \approx 5$). 
According to this new calibration, a multiplicative factor of $ \approx 
$ 0.75 must be applied to the COBE/IRAS dust emission reddening maps.    
}

\titlerunning{On the calibration of the COBE/IRAS dust emission reddening maps}
\authorrunning{Dutra et al.}

\maketitle
 
\section{Introduction} 

Galactic interstellar reddening is one of the observational limitations often challenging 
astronomers. Accurate maps of Galactic reddening are of 
crucial importance for a number of applications, such as measuring distances and mapping the 
peculiar velocity field in the nearby universe. Because of the patchy dust distribution 
in our Galaxy, it is very difficult to establish 
a general Galactic extinction law. 
However, various projects were undertaken in this direction. Sandage (1973)
 and 
de Vaucouleurs et al. (1976) modeled the interstellar extinction distribution as a 
function of the Galactic
 coordinates ($\ell$, $b$). These general extinction laws were mainly 
used in extragalactic studies
 and have been adopted in the Shapley-Ames (Sandage \& 
Tamman 1981) and Second Reference (de Vaucouleurs et al. 1976) galaxy catalogues, respectively.
 
Burstein \& Heiles (1978, 1982) found relationships between galaxy counts, HI column densities 
and
  $E(B-V)$ reddening values. Additionally, they worked out reddening maps covering a 
great part of the sky, excepting the latitudes lower than $|b|<10^{\circ}$.
 Since the dust 
grains lie in the interstellar medium associated with the HI gas (Reach et al. 1998), 
Burstein \& Heiles' model reproduced 
the irregularities of the dust distribution more 
precisely. This is one of the reasons why Burstein \& Heiles maps have been widely used 
in the literature. More recently, Schlegel et al. (1998, hereafter SFD98) 
provided a new estimator of Galactic reddening by means of a full-sky 100 $\mu$m IRAS/ISSA map, 
which was converted to dust column density by using a dust colour temperature map (17 K to 21 K) 
derived from 100 and 240 $\mu$m COBE/DIRBE dust emission maps. We will refer to these reddening maps  from the  IRAS/ISSA and COBE/DIRBE experiments
by COBE/IRAS dust emission reddening maps, following Chen et al. 1999. The calibration of 
the relationship between the dust column density or opacity $\tau _{FIR}$ and SFD98's dust emission reddening 
$E(B-V)_{FIR}$ was made using a sample of early-type galaxies with an uncertainty of about 
16\% in their reddening values, up to $E(B-V)=0.15$. The COBE/IRAS dust emission reddening maps cover the whole sky 
with a 
resolution of 6.1$^{\prime}$.  They show a
 good agreement with Burstein \& Heiles' maps at intermediate and high Galactic
latitudes. However, SFD98 did not test their maps towards low Galactic latitudes and in the 
directions of the Magellanic Clouds. 
Given their characteristics, SFD98's maps supersede those of 
Burstein \& Heiles and different authors have tested the $E(B-V)_{FIR}$ values with 
independent reddening estimates derived from the stellar content of Galactic objects. 
 Hudson (1999) compared the reddening predictions of these maps with the 
$E(B-V)$ values estimated for 50 distant globular clusters with $|b|>10^{\circ}$
 and 
distances perpendicular to the Galactic plane $|Z|>3$ kpc, as well as with those of 86 
RR Lyrae stars from the sample of Burstein \& Heiles (1978). The reddening comparisons were 
carried out up to $E(B-V)\approx$ 0.30 and resulted in reddening differences of $\delta 
E(B-V)=$ -0.008 and -0.016 for the two samples, respectively.

On the other hand, 
Dutra \& Bica (2000) compared $E(B-V)$ reddening values from the stellar content 
of 103 old open clusters and 147 Galactic globular clusters with those derived from 
$E(B-V)_{FIR}$ maps. They obtained comparable reddening values between the two 
procedures for star clusters at $|b|>20^{\circ}$, in agreement with the fact that 
most of these clusters are located beyond the disk dust layer. However, Dutra \& Bica (2000) 
found significant differences for clusters at low Galactic latitudes located or projected in the disk 
dust layer. For low Galactic latitude star clusters, the differences are due to the 
background dust contribution, since the heated dust in the plane and towards the Galactic 
centre contributes only to the $E(B-V)_{FIR}$ values. Considering 131 globular clusters 
with $|b|>2.5^{\circ}$ and $|Z|> 100$ pc (assumed dust scale height), Chen et al. (1999) 
concluded that SFD98's reddening maps overestimated the visual absorption by a mean 
factor of 1.16. Arce \& Goodman (1999) analyzed
 the interstellar reddening in the Taurus 
dark cloud complex ($b\approx -15^{\circ}$)  by comparing $E(B-V)_{FIR}$ 
reddening values with those derived from four other methods and concluded that SFD98's 
reddening values overestimated the extinction by a factor of 1.3-1.5 in regions of smooth 
extinction with $A_{V}$ $>$ 0.5. Dutra et al. (2002) built $A_K$ extinction maps 
using 2MASS data for
 two low extinction regions (windows) in the inner Bulge. The 
comparison of extinction values within a radius of one degree of window W359.4-3.1 (at 
$b=3.1^{\circ}$, and mean $E(B-V)=$ 0.8) indicated that
 the dust emission extinction values are 
overestimated by a factor of 1.45 with respect to those derived from 2MASS photometry. Also using 2MASS data, 
Dutra et al. (2003) built an $A_K$ extinction map within ten degrees of
 the Galactic centre, 
finding that for $3^{\circ}<|b|<5^{\circ}$, the discrepancy between reddening values 
derived from the dust emission and 2MASS photometry can be explained by a calibration factor of 
1.31.
 
Dutra et al. (2001, hereafter Paper I) obtained spectra from the nuclear region of galaxies 
behind and around the Magellanic Clouds in order to estimate
 the reddening in their lines of sight. 
They derived the reddening in each line-of-sight by matching the continuum distribution of the galaxy's spectrum 
with that of a reddening-free template
 with similar stellar population. For galaxies in the surroundings of the Magellanic Clouds, they concluded that the spectroscopic and dust emission reddening values agree well. They also detected the effect of the internal reddening of the Magellanic Clouds for galaxies 
behind them.
 
 From the studies mentioned above, we can  infer that the dust emission reddening 
maps - with their present reddening-dust emission calibration - need an additional 
calibration factor to reproduce reddening values higher than $E(B-V)=$ 0.3.
 In the current work, we use 
the spectroscopic method described in Paper I to derive independent foreground reddening estimates for
early-type galaxies in lines of sight  at low Galactic latitudes ($b<25^{\circ}$), in order to compare 
these reddening values with those derived from dust emission. By
 using the present sample 
of galaxies, which provides reddening estimates for a wide sky coverage and by 
combining them to results cited above, we aim to extend the calibration 
of the dust emission 
reddening maps  to regions with $|b|>4^{\circ}$ and more heavily reddened.

In Sect. 2 we present the sample of the observed early-type galaxies at low Galactic 
latitudes.
 In Sect. 3 we describe the observations and reductions, whereas in Sect. 
4
 we compare the present galaxy spectra with those of red stellar
 population templates obtained 
in Paper I  to derive the reddening values. We also discuss   
some discrepancies appearing between the spectroscopic reddening values and those
 derived from the 100 $\mu $m dust emission 
reddening maps. In Sect. 5 we provide a new reddening calibration for the COBE/IRAS dust emission reddening maps 
using the galaxy sample and extend it  using literature data. In 
Sect. 6 we test the new calibration using the intrinsic $(B-V)_0$ colour $\times Mg_2$ index relation for 311 galaxies.
 Concluding remarks are given in Sect. 7.

\section{The sample} 

\begin{figure}
\resizebox{\hsize}{!}{\includegraphics{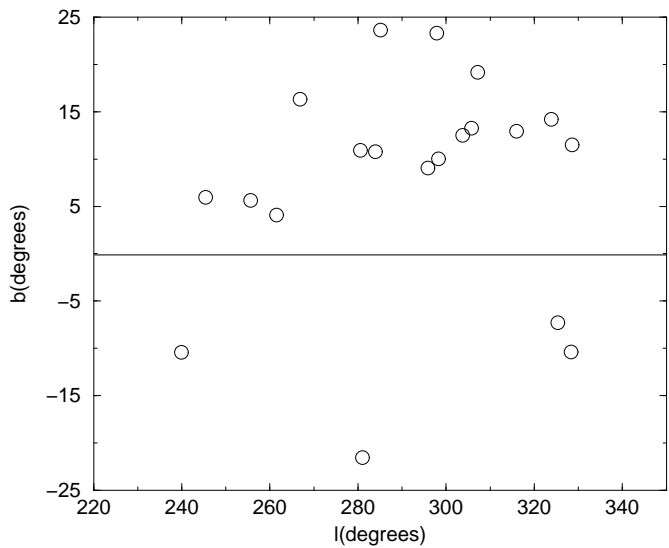}}
\caption{Angular distribution of the observed sample of low Galactic latitude galaxies.}
\label{sample}
\end{figure}

We used the LEDA database ({\rm http://leda.univ-lyon1.fr}) as a starting point to 
select early-type galaxies with Galactic latitude
 $|b|<25^{\circ}$ and total blue 
magnitude $B_T<15$. Figure 1 shows the angular distribution of the observed galaxies. 
The present 
sample and its properties are given in Table 1, which includes the following columns: 
(1) designation, (2)
 and (3) J2000 equatorial coordinates, (4) and (5) Galactic coordinates, (6) 
total blue magnitude $B_T$, (7) exposure time, (8) radial velocity
 measurement, (9) 
LEDA and/or NED radial velocity, and (10) LEDA/NED  morphological type.

\begin{table*}
\tiny 
\caption{\scriptsize 
The sample of observed galaxies.} 
\begin{tabular}{lccccccccl} 
\hline\hline 
Object&RA(2000)&Dec(2000)&$\ell$&$b$&$B_t$&Exp&$V$&$V_{lit}$&Type\\
&h:m:s~~~&$^{\circ}$:$^{\prime}$~:$^{\prime\prime}~$&($^{\circ}$)&($^{\circ}$)&&(sec)&(km/s)&(km/s)&\\
\hline
&&&&CASLEO 2.15m sample&&&&\\
\hline
PGC32955&10:57:48&-47:50:36&283.96&10.80&14.49&3$\times$1200&5761&--&E-S0\\
ESO137-45&16:51:03&-60:48:30&328.34&-10.39&13.19&3$\times$1200&3450&3351&E\\
ESO171-8&12:04:46&-53:10:27&295.88&9.06&14.19&3$\times$1200&4544&4423&E\\
ESO218-2&12:21:09&-52:35:09&298.27&10.01&13.66&2$\times$1200&4520&4280&E\\
ESO221-26&14:08:24&-47:58:13&316.01&12.94&12.10&2$\times$1200&1517&2788&E\\
ESO264-31&10:40:33&-46:11:29&280.53&10.91&13.91&3$\times$1200&6605&6724&E\\
ESO269-72&13:13:34&-43:31:35&307.18&19.16&13.98&3$\times$1200&2912&3084&E\\
ESO273-2&14:46:31&-43:57:10&323.87&14.19&14.19&2$\times$1200&5200&--&E-S0\\
ESO274-6&15:16:10&-44:00:50&328.61&11.52&14.45&3$\times$1200&4942&4458&E\\
ESO314-2&08:57:29&-39:16:09&261.50&4.10&12.91&1200&1297&982&E\\
IC3896&12:56:44&-50:20:43&303.80&12.52&12.18&2$\times$1200&2148&2110&E\\
NGC2663&08:45:08&-33:47:40&255.67&5.65&11.99&2$\times$1200&2243&2122&E\\
ESO137-8&16:15:46&-60:55:07&325.34&-7.28&13.21&3$\times$1200&4110&3833&E\\
\hline
&&&&ESO 1.52m sample&&&&\\
\hline
IC2311&08:18:46&-25:22:14&245.44&5.95&12.50&2$\times$1800&1960&1844&E\\
NGC3087&09:59:09&-34:13:30&266.89&16.32&12.46&1800&2680&2626&E\\
NGC3706&11:29:45&-36:23:27&285.08&23.63&12.33&2$\times$900&3100&2978&E-S0\\
NGC2325&07:02:40&-28:41:50&239.96&-10.41&12.30&2$\times$1800&2050&2164&E\\
NGC4976&13:08:38&-49:30:17&305.80&13.27&10.96&2$\times$1200&1600&1444&E\\
IC3370&12:27:37&-39:20:16&297.92&23.30&12.03&2$\times$1800&3070&2935&E\\
NGC2434&07:34:52&-69:17:02&281.00&-21.54&12.36&2$\times$1800&1550&1400&E\\
\hline
\end{tabular} 
\end{table*}

\section{Observations and reductions}

\begin{table*}
\tiny 
\caption{\scriptsize 
Instrument and spectra general characteristics.} 
\begin{tabular}{lccccccccc} 
\hline\hline 
Observatory/tel&spectrograph&size CCD&pixel size&grating&dispersion&spectral coverage&width 
slit&resolution&lenght slit\\
&&(pixels)&($\mu m^2$)&(grooves mm$^{-1}$)&(\AA/pixel)&(\AA)&($^{\prime\prime})$&(\AA)&($^{\prime}$)\\
\hline
ESO 1.52m&B\&C&2688$\times$ 512&15$\times$ 15&300&3.72&3600-10100&3.5&18&4.1\\
CASLEO 2.15m&B\&C&1024$\times$1024&24$\times$ 24&300&3.43&3600-6800&4.0&14&4.7\\
\hline
\end{tabular} 
\end{table*}
 
The spectra were collected with  the 1.52-m telescope at the European Southern 
Observatory (ESO, La Silla, Chile) on 11-14 January 2002 and 6-7 December 2002, and the 2.15-m telescope at the 
Complejo Astron\'omico El Leoncito (CASLEO, San Juan, Argentina) on 10-12 May 2002.  In both 
telescopes we employed a CCD camera 
attached to a Boller \& Chivens spectrograph. 
At least two
 exposures of each object were taken in order to correct for cosmic 
rays. The exposure 
times are given in Table 1, while Table 2 summarizes the instrumental and general 
characteristics of the spectra obtained in the two Observatories. The standard stars 
EG21, 
LTT2415,  and LTT3864 (Baldwin \& Stone 1984) were observed for flux 
calibrations. He-Ar lamp 
exposures were taken following that of 
the object or standard star for wavelength calibrations. 
The slit was set in the E-W direction and 
its length projected on the chip (Table 2) provided a 
wide range of 
pixel rows for background subtractions.

The reductions were carried out using the IRAF package following standard procedures. 
The 
spectra were bias corrected, flat-fielded and extracted along the slit
 according to the 
dimensions of each galaxy's nuclear region. Typical
 extractions were 
$\approx$ 8-10$^{\prime\prime}$. Afterwards, they were calibrated in wavelength and 
flux units using a He-Ar lamp and standard star spectra, respectively. Corrections for 
atmospheric extinction were applied, adopting the standard
 mean extinction curves from ESO/La 
Silla and CASLEO (Minniti et al. 1989).
   
Since the spectral resolution was chosen for stellar population purposes, it is not 
ideal for radial velocity measurements. We measured in any case the radial velocities using the 
absorption bands/lines centred at the
G band (4301 \AA), H$_{\beta}$ (4861 \AA), Mg I (5175 \AA) and 
Na I (5890 \AA).
 The measured velocities and those from the LEDA and NED extragalactic databases 
are consistent (Table 1). In one case, ESO\,221-26,
 the LEDA velocity (Table 1) appears to be significantly different from that of 
present observation. The measured 
velocities were used to bring the galaxy spectra to the rest frame, which is
 necessary for 
the subsequent stellar population analysis.

Figures 2, 3 and 4 show the resulting rest-frame flux 
calibrated spectra for the observed 
galaxies at CASLEO 2.15-m and ESO 1.52-m telescopes.
Note that although the spectra 
taken with the ESO 1.52-m telescope yield a larger spectral coverage (Table 2), we adopted 
the range (3600 - 6800\AA) for all the spectra.

\begin{figure}
\resizebox{\hsize}{!}{\includegraphics{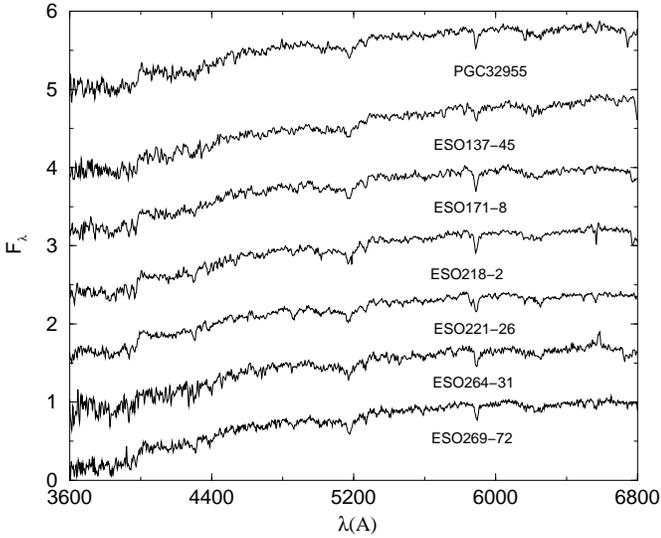}}
\caption{Rest-frame spectra of galaxies observed at CASLEO. Spectra are in relative 
$F_\lambda$ units normalized at 5870 \AA. Constants have been added to the spectra 
for the sake of clarity, except for the bottom one.}
\label{sample}
\end{figure}

\begin{figure}
\resizebox{\hsize}{!}{\includegraphics{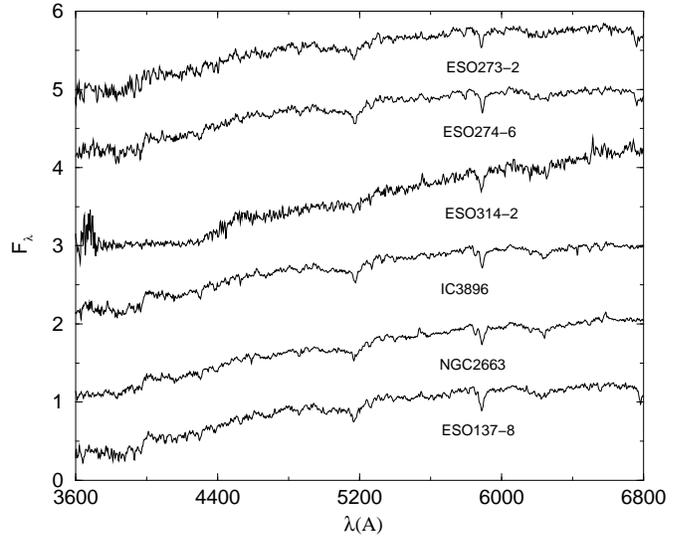}}
\caption{Rest-frame spectra of additional galaxies observed at CASLEO. Units are as in Fig. 2.}
\label{sample}
\end{figure}

\begin{figure}
\resizebox{\hsize}{!}{\includegraphics{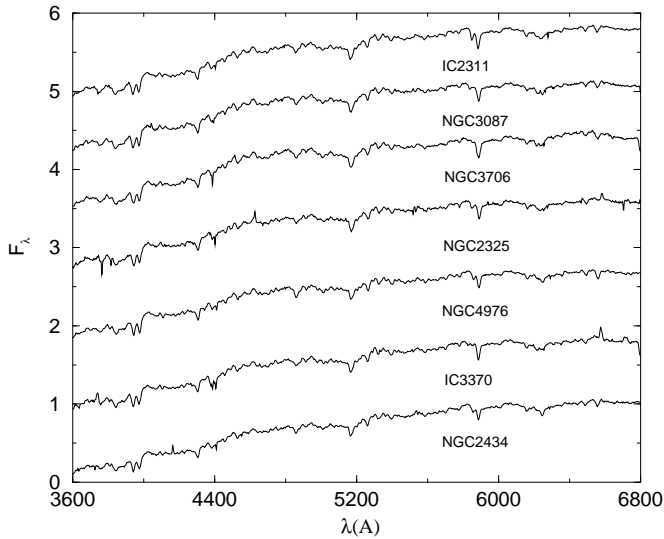}}
\caption{Rest-frame spectra of galaxies observed at ESO (La Silla). Units are as in Fig. 2.}
\label{sample}
\end{figure}

\begin{table*} 
\tiny
\caption{\scriptsize 
Equivalent widths for strong absorption features in the templates and individual galaxy spectra}
\centerline{ 
\begin{tabular}{lccccccccc} 
\hline\hline 
Object & K & H$_{\delta} $ & CN & G & H$_{\gamma} $ & H$_{\beta} $ & MgI & NaI 
& H$_{\alpha} $ \\
Windows & 3908-3952 & 4082-4124 & 4150-4214 & 4284-4318 & 4318-4364 & 
4846-4884 & 5156-5196 & 5880-5914 & 6540-6586   \\
\hline
&&&&&Red Population Templates&&&&\\
\hline
T1&18.6&6.1&15.0&9.6&4.3&3.4&11.3&7.0&0.2\\
T23&17.1&4.3&10.2&9.2&4.4&3.5&8.7&4.3&0.8 \\
\hline
&&&&&CASLEO 2.15m sample&&&&\\
\hline
PGC32955&20.4&4.7&18.6&11.7&9.7&1.1&9.7&5.8&2.2\\
ESO137-45&22.5&1.6&11.1&8.9&7.7&3.9&10.2&5.1&1.0\\
ESO171-8&19.1&5.0&15.3&8.9&7.1&3.8&10.6&7.1&1.4\\
ESO218-2&19.0&6.9&16.4&10.7&5.3&3.7&11.8&6.2&2.5\\
ESO221-26&18.4&6.8&13.1&9.3&5.6&4.9&9.0&5.5&2.3\\
ESO264-31&17.6&5.1&13.1&11.2&7.2&3.7&9.8&6.1&--\\
ESO269-72&20.1&6.3&14.9&10.4&5.9&3.8&10.1&4.6&2.7\\
ESO273-2&16.9&8.2&16.6&9.0&4.8&3.6&9.5&5.3&1.3\\
ESO274-6&20.0&6.9&15.8&10.1&5.8&3.6&10.9&6.6&1.4\\
ESO314-2&34.1&30.3&57.3&26.1&21.7&5.2&10.3&7.0&1.8\\
IC3896&20.9&7.4&18.5&10.6&6.3&4.4&11.9&6.5&2.2\\
NGC2663&17.5&5.9&18.5&10.3&7.6&4.9&10.7&6.7&0.4\\
ESO137-8&22.3&9.4&18.0&10.1&7.4&3.8&11.4&7.2&2.5\\
\hline
&&&&&ESO 1.52m sample&&&&\\
\hline
IC2311&17.3&6.9&15.7&11.1&7.4&4.6&9.6&5.1&1.6\\
NGC3087&16.7&6.6&15.3&11.6&7.2&4.2&9.7&4.8&1.6\\
NGC3706&17.4&7.2&16.5&11.6&7.4&3.9&10.1&6.0&2.0\\
NGC2325&17.2&5.1&14.1&10.4&6.3&3.6&9.7&5.1&0.1\\
NGC4976&15.0&6.8&13.4&10.5&8.3&5.6&8.5&4.6&2.5\\
IC3370&14.7&4.9&12.7&10.2&6.4&4.6&9.7&5.0&--\\
NGC2434&15.8&5.5&11.5&10.8&7.2&4.7&9.0&4.4&1.7\\
\hline
\end{tabular}}
\end{table*}

\section{Reddening Analysis}  

All the observed spectra (Figs. 2, 3 and 4) correspond to the nuclear region of 
early-type galaxies and, therefore, they are characterized by a red 
stellar population.
 Bica (1988) studied the stellar populations of the nuclear regions 
of early and late-type galaxies by means of their integrated spectra. 
Red stellar population galaxy nuclei spectra are ideal 
as Galactic reddening probes since the spectral distribution is essentially
 insensitive to age 
variations of the components and presents a small 
dependence on metallicity. He  grouped the spectra in order to form
 high 
signal-to-noise red and blue stellar population templates, which are corrected for 
reddening following Sandage's law (1973). These templates represent the most 
frequent types of stellar populations found in normal galaxy nuclei.

In Paper I we defined 
new red stellar population templates, T1 and T23, based on those of Bica (1988) and 
new observations of early-type galaxies in the South Polar Cap. These templates are 
reddening-corrected and formed by galaxies with dust emission reddening $E(B-V)_{FIR}$ 
$<$ 0.02. 
We adopted the T1 and T23 templates in the present study to perform the match 
between the continuum distribution of these spectra and those from
 the observed 
galaxies for the purpose of estimating the reddening in their lines of sight.

Internal reddening in early type galaxies might be a source of uncertainty in the present method. 
Van Dokkum \& Franx (1995)  found high opacities in cores of ellipticals in a scale of 1-2$^{\prime\prime}$ using HST. Ferrari et al. (1999) showed the presence of internal dust in 75\% of 22 observed elliptical galaxies. By integrating in larger areas (10-20$^{\prime\prime}$) they derived a typical reddening value of $E(B-V)=0.01$, thus within uncertainties of the present determinations (typically $\epsilon _{E(B-V)}=0.02$, Paper I). Therefore the internal reddening variations in ellipticals are not expected to affect the present results significantly, since our apertures are relatively large (Sect. 3). 
 
For the spectral comparisons, we need to determine which template has a stellar 
population
 which most closely resembles that of the observed galaxy. This estimation is 
made by comparing the equivalent widths (Ws) 
of both spectra. Then, we employ 
Ws of strong absorption features from the sample galaxies and templates' spectra. 
We use the following metal
 features: K CaII (3933 \AA), CN (4182 \AA), G band 
(4301 \AA), MgI (5175 \AA) and NaI (5890 \AA), together with four
 Balmer lines: 
H$_{\delta}$ (4101 \AA), H$_{\gamma}$ (4340 \AA), H$_{\beta}$ (4861 \AA) and H$_{\alpha}$ 
(6563 \AA). Table 3 shows W values for templates and
 individual galaxies measured with 
continuum tracings and feature
 windows following Bica \& Alloin (1986) and Bica et al. (1994). 
Typical W errors
 are $\approx$ 5 \% and depend mostly on uncertainties in the
 continuum positioning. 

For each galaxy the spectroscopic reddening value $E(B-V)_{SPEC}$ was obtained by fitting 
the observed galaxy spectrum to that of the corresponding template with
 similar Ws, 
by applying continuum corrections according to the Seaton's (1979) Galactic extinction law.
 
Fig. 5 illustrates the reddening determination for NGC\,4976 (upper panel) and 
NGC\,2663 (lower panel),
 of stellar populations similar to those of the  templates 
T23 and T1, respectively. The reddening effect in the 
observed spectra are important.
 Table 4 provides the associated template, as well as the 
spectroscopic and dust emission reddening value for each galaxy.

Fig. 6 shows the comparison between these two reddening estimates using the present data 
and those from Paper I for
 34 early-type galaxies at intermediate latitudes, surroundings of the Magellanic 
Clouds and South Polar Cap. A qualitative analysis shows that the two reddening estimates are 
consistent up to $E(B-V)< 0.25$, but beyond this limit
the two available points indicate higher $E(B-V)_{FIR}$
 values. Fig. 6 suggests a discrepancy between
spectroscopic and higher dust emission reddening values, which  has also been detected  
in other reddening comparisons
in the literature, as mentioned in Sect. 1. The spectroscopic reddening
determination method is a promising technique to understand this discrepancy
 at high reddening values regime with a homogeneous way of expanding the sample
 to fainter galaxies and extending the analysis towards the infrared domain.

\begin{figure}
\resizebox{\hsize}{!}{\includegraphics{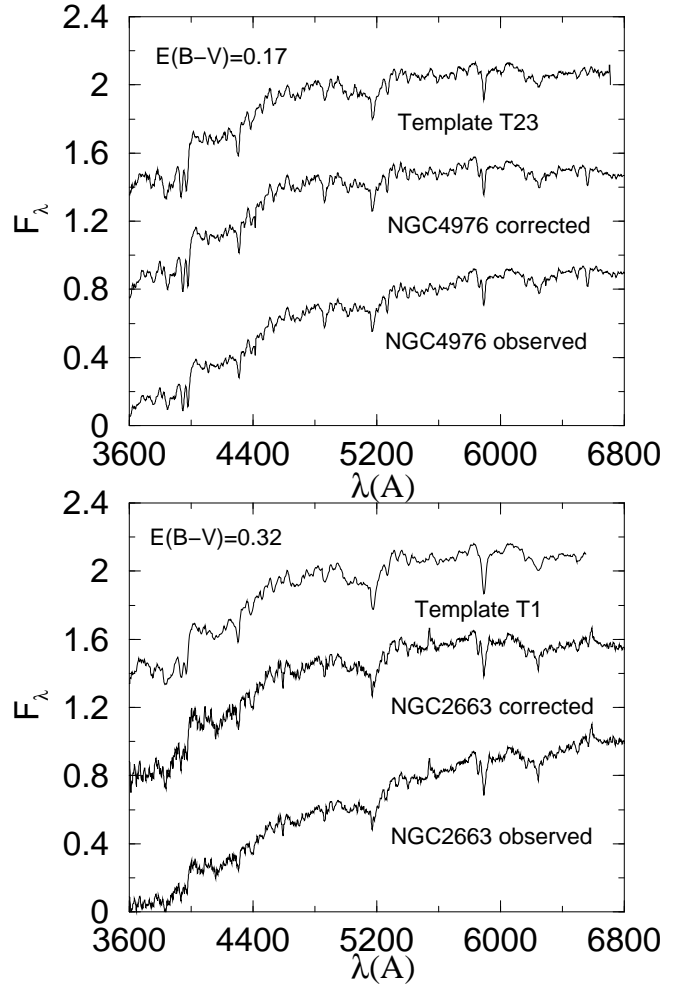}}
\caption{Upper panel: reddening determination for NGC\,4976. Lower panel: reddening
determination for NGC\,2663.}
\label{sample}
\end{figure}

\begin{table} 
\caption{\scriptsize 
Spectroscopic and dust emission reddening values for observed galaxies from 
Bica (1988) and South Polar Cap samples.} 
\begin{tabular}{lccccc} 
\hline\hline 
Object&Template&$E(B-V)$&$E(B-V)_{FIR}$\\
\hline
PGC32955&T1&0.17&0.21\\
ESO137-45&T1&0.25&0.24\\
ESO171-8&T1&0.14&0.16\\
ESO218-2&T1&0.15&0.17\\
ESO221-26&T23&0.16&0.22\\
ESO264-31&T1&0.14&0.16\\
ESO269-72&T1&0.14&0.16\\
ESO273-2&T1&0.14&0.17\\
ESO274-6&T1&0.15&0.17\\
ESO314-2&T1&0.65&0.76\\
IC3896&T1&0.20&0.21\\
NGC2663&T1&0.32&0.36\\
ESO137-8&T1&0.20&0.19\\
IC2311&T1&0.14&0.14\\
NGC3087&T1&0.10&0.11\\
NGC3706&T1&0.10&0.09\\
NGC2325&T1&0.12&0.12\\
NGC4976&T23&0.17&0.19\\
IC3370&T1&0.14&0.09\\
NGC2434&T23&0.25&0.25\\
\hline
\end{tabular}
\end{table}

\begin{figure}
\resizebox{\hsize}{!}{\includegraphics{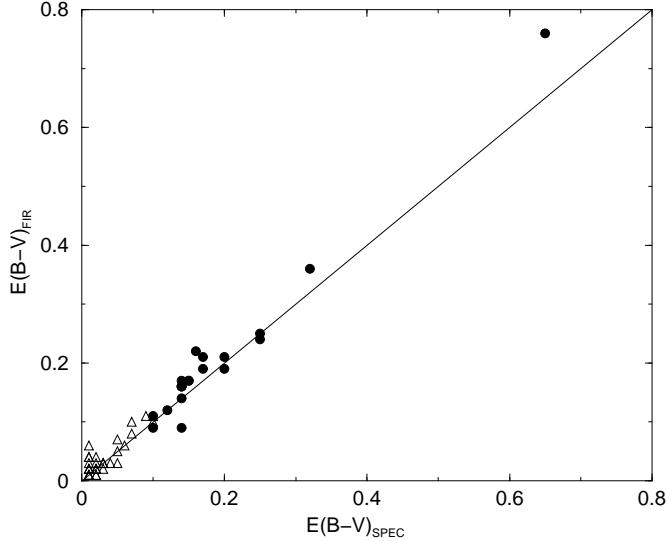}}
\caption{Comparison between $E(B-V)$ reddening values derived from galaxy spectra and those from 100 $\mu$m dust emission.
 The present data are indicated by filled 
circles, while Paper I's data are indicated by open triangles.}
\label{sample}
\end{figure}

\section{Recalibrating COBE/IRAS dust emission reddening maps.}

\begin{figure}
\resizebox{\hsize}{!}{\includegraphics{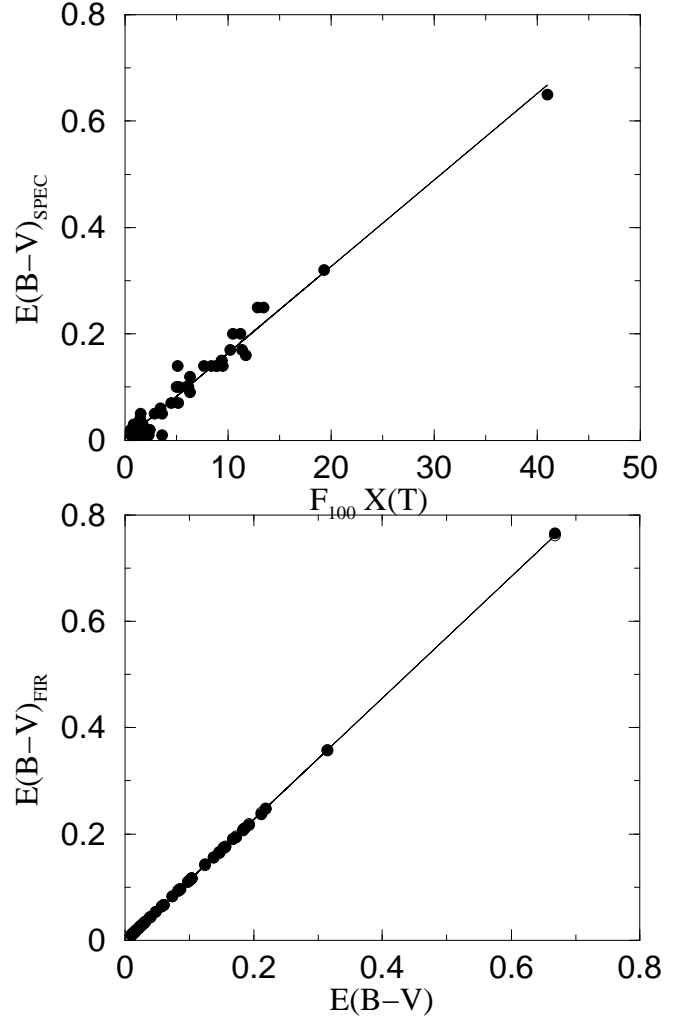}}
\caption{Comparison of reddening determinations: (i) upper panel, 
spectroscopic $E(B-V)_{SPEC}$ values {\it versus} SFD98's opacities $F_{100} X(T)$; (ii) lower panel, SFD98's dust emission reddening 
$E(B-V)_{FIR}$ values {\it versus} spectroscopic $E(B-V)$ colour excesses derived from Eq. (2).}
\label{sample}
\end{figure}

\begin{figure}
\resizebox{\hsize}{!}{\includegraphics{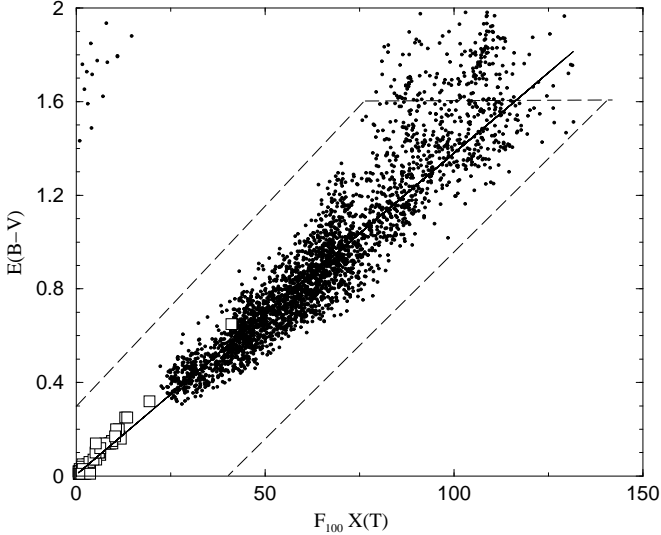}}
\caption{Independent reddening values $E(B-V)$ {\it versus} SFD98's opacities $F_{100} X(T)$ for galaxy sample (open squares) and 3026 Bulge direction (filled circles).}
\label{sample}
\end{figure}

The transformation of opacity $\tau _{FIR}$ to
a reddening $E(B-V)_{FIR}$ map carried out by SFD98 uses the correlation between the intrinsic
$(B-V)_0$ colour of elliptical galaxies and their $Mg_2$ index. 
The $Mg_2$ index described by Faber et al. (1989) has a tight correlation with $(B-V)_0$,
which in turn can be used to obtain accurate reddening values.
SFD98 used 389 elliptical galaxies from Faber et al. (1989) to compute a linear regression of 
reddening-corrected $(B-V)_0$ values against $Mg_{2}$ with residuals 
$\delta (B-V)_0$. They computed the Spearman rank correlation coefficient of 
$\delta (B-V)_0$ versus the $E(B-V)_{FIR}$ values, arguing that a 
good dust map will have no residual correlation. SFD98 obtained the 
following calibration factor:
 
 \begin{equation}
 p=\frac{E(B-V)}{F_{100}X(T)}=0.0184 \pm 0.0014,
 \end{equation}
 
 \noindent where $F_{100}$ is the 100$\mu$m IRAS flux expressed in MJy sr$^{-1}$ corrected
  from zodiacal emission, and X(T) is the temperature correction factor to
 the fixed dust grain temperature (18.2 K) derived
 from the 100/240$\mu$m COBE/DIRBE maps. The product $F_{100}X(T)$ is defined as the FIR opacity $\tau _{FIR}$.

Arce \& Goodman (1999) found that SFD98's reddening maps overestimated
the reddening in the Taurus dark cloud complex by a factor  of 1.3-1.5 
in regions of high extinction ($A_V>0.5$). They attributed this discrepancy
to the fact that in 
the sample of 389 elliptical galaxies used to calculate a conversion from 
dust column density to $E(B-V)$, 90\% of the galaxies have low-reddening 
($E(B-V)_{FIR} <$ 0.1) and very few have high-reddening 
($E(B-V)_{FIR} >$ 0.15) values. The lack of galaxies in 
high-reddening regions results in an inaccuracy in the conversion between 
dust column and reddening for lines of sight with $E(B-V)_{FIR}>0.15$.

 We use the present reddening estimates
derived with the spectroscopic method to perform a new calibration of SFD98's 
opacities derived from 100$\mu$m emission and temperature correction maps.
Fig. 7's upper panel shows the comparison between spectroscopic reddening estimates and 
the corresponding SFD98's opacities. A linear 
regression to the data yields the following equation:

\begin{equation}
E(B-V) = 0.016 F_{100} X(T) + 0.0008
\end{equation}

\noindent which includes a calibration factor slightly lower than that of 
Eq. (1). Fig. 7's lower panel shows the comparison 
between SFD98's reddening estimates and those derived from Eq. 
(2) using the opacities $\tau _{FIR}$. A linear regression to 
the data yields now:

\begin{equation}
E(B-V)_{FIR} = 1.1412 E(B-V) - 0.0015
\end{equation}

\noindent which indicates that 
$E(B-V)_{FIR}$ values overestimate reddening values by a factor of $\approx$ 1.14 
 with respect to the present calibration for reddening values up to $E(B-V) \approx 0.7$. 
  This result is in good agreement with that of Chen et al. (1999) who
 found that SFD98's reddening values overestimate the visual absorption
  by a factor of 1.16 for a sample of globular and open clusters with $|b|>2.5^{\circ}$,
 most of them with absorption $A_V<3$ ($E(B-V)<1$).

In order to extend further the opacity-reddening calibration to higher $E(B-V)$ values, 
the K-band extinction $A_{K}$  map built by 
Dutra et al. (2003) within ten degrees of the Galactic center was used, together with the present reddening estimates.

Dutra et al. (2003) determined the $A_K$ extinction from $JHK_s$ infrared 2MASS photometry of Bulge red giants,
 fitting the upper giant branch of ($K_s$, $J-K_s$) colour-magnitude diagrams to a dereddened upper giant branch mean locus. These Bulge/Disk directions were divided into cells and about 6 million stars were used as reddening probes.
 Adopting $R_V = A_{V} / E(B - V)$ = 3.1 and $A_K/A_V=0.112$ (Cardelli et al. 1989), we transformed $A_K$ estimates to $E(B-V)$ ones for 3026 cells with 4$\times$4 arcmin$^{2}$ in size,
  located at Galactic coordinates $|\ell|<5^{\circ}$ and $4^{\circ}<|b|<5^{\circ}$, in a region where the dust background contribution to the 100$\mu$m flux (less than 5\%) is minimized
  and 
$A_K > 2.5 \sigma _{i}$, where $\sigma _{i}$ is the uncertainty in the $A_K$ determination. The latter constraint is to warrant accuracy for the $A_K$ estimates.
Fig. 8 shows for the galaxy sample and 3026 Bulge directions 
the comparison between independent reddening estimates and the SFD98's opacities. 
A linear regression fit within the polygon region (Fig. 8) yields the following new 
calibration:

\begin{equation}
E(B-V)_{CAL} = 0.0137 F_{100} X(T) + 0.0064,
\end{equation}

\noindent which extends up to $E(B-V)=1.6$ (or $A_V \approx 5$).

Fig. 9's panels (a) and (b) show for the galaxy sample and 3026 Bulge directions the comparisons between independent reddening estimates and
 those derived from SFD98 $E(B-V)_{FIR}$ and $E(B-V)_{CAL}$, respectively. In panel (a), 
 the $E(B-V)_{FIR}$ values appear to be systematically larger than the $E(B-V)$ ones, while
 panel (b) shows a good agreement between $E(B-V)_{CAL}$ and  $E(B-V)$, except for the points having $E(B-V)_{CAL}<0.4$ and $1.4<E(B-V)<2.7$. 
The latter points correspond to those with $E(B-V)_{FIR}>2.4$  and 
the same $E(B-V)$ range. In addition, Eq. (4) is not precise for higher reddening values of $E(B-V)>1.6$, where the
  slope of the $E(B-V)_{CAL} \times F_{100} X(T)$ correlation has an abrupt variation (Fig. 8). In Fig. 9's panel (c) 
 we present histograms of the relative differences $\chi _{CAL}$ and $\chi _{FIR}$, between $E(B-V)_{CAL}$ and independent $E(B-V)$ measurements
 and $E(B-V)_{FIR}$ and $E(B-V)$ values, respectively. The relative differences
  $\chi _{CAL}$  and $\chi _{FIR}$ are defined as:

\begin{eqnarray}
\left\{ \begin{array}{ll}
\chi _{CAL}= 100 \times \frac {\sqrt {(E(B-V)_{CAL} - E(B-V))^{2}}}{E(B-V)} \\
\\ 
\chi _{FIR}= 100 \times \frac {\sqrt {(E(B-V)_{FIR} - E(B-V))^{2}}}{E(B-V)} \\
\end{array}
\right.
\end{eqnarray}

As shown in Fig. 9's panel (c), 90\% of the points in the new 
  calibration have a relative difference $\chi _{CAL}<20\%$, while the $\chi _{FIR}$ distribution  has a peak at $\approx$ 35\%.
   This result shows that the present calibration  proves to be an improvement  in the studied reddening range. 
A comparison between present and SFD98's reddening estimates is 
shown in Fig. 9's panel (d). A linear regression fit within the polygon 
region (3044 points) leads to the following relation:

\begin{equation}
E(B-V)_{CAL} = 0.7480 E(B-V)_{FIR} + 0.0056
\end{equation}

Eq. (6) indicates that besides better reproducing the 
present data, our calibration could additionally provide good agreement with 
independent reddening measurements made by Arce \& Goodman (1999) in the 
Taurus dark cloud.
 
\begin{figure}
\resizebox{\hsize}{!}{\includegraphics{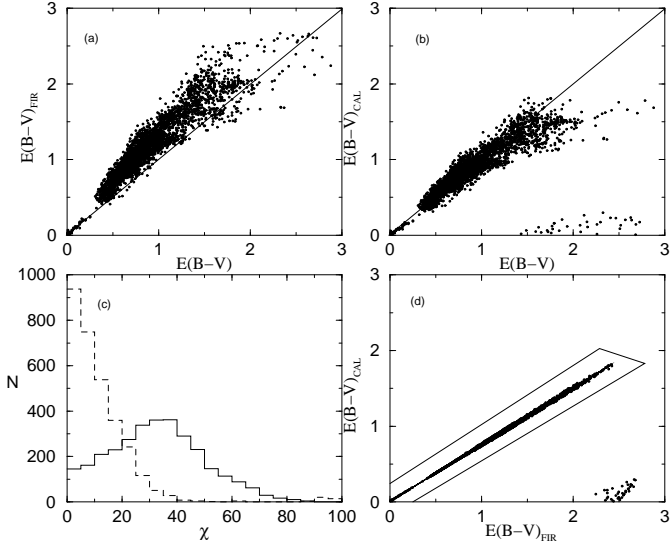}}
\caption{
Panels: (a) $E(B-V)_{FIR}$ {\it versus} $E(B-V)$, (b) 
$E(B-V)_{CAL}$ {\it versus} $E(B-V)$, (c) histograms of the relative differences 
$\chi _{CAL} $ (dashed line) and $\chi _{FIR} $ (solid line), (d) 
$E(B-V)_{CAL}$ {\it versus} $E(B-V)_{FIR}$.}
\label{sample}
\end{figure}

\section{Testing the FIR opacity-reddening calibration}

Hudson (1999) examined the large-scale systematic errors, which vary as a function of position on the sky, in Burstein \& Heiles and SFD98 
reddening maps. He determined for 311 early-type galaxies a relation $Mg_2$ index and  intrinsic $(B-V)_0$ colour. 
We used the data by Hudson (1999) available at CDS to study the
correlation between the $Mg_2$ index and $(B-V)_0$ colour; for such, 
we dereddened the observed $(B-V)$ colour using
 SFD98's reddening estimates and those from the present calibration. Fig. 10 shows in panels (a) and (c) the $(B-V)_0$-$Mg_2$ 
 relation, once the observed $(B-V)$ colours were 
corrected for reddening using SFD98's $E(B-V)_{FIR}$ and the present calibration $E(B-V)_{CAL}$, respectively. Panels (b) and (d) show
 the residuals from a linear fit to the $(B-V)_0$-$Mg_2$ as a function of the respective reddening estimates.
 Comparing the residual plots, we note that for the same sample, $E(B-V)_{FIR}$ values are higher
  than $E(B-V)_{CAL}$ ones. The residual plot from the linear fit in (c) as a function of $E(B-V)_{CAL}$ shows a more symmetric distribution than 
  that for $E(B-V)_{FIR}$ values, indicating that the former is less biased than the latter, mainly for higher reddening values. Blakeslee et al. (2001) 
  performed the same test for 200 galaxies using a relation between the intrinsic $(V-I)_0$ colour and the $Mg_2$ index.
   They concluded that SFD98 reddening values overestimate 
  by 5-10\% the interstellar reddening in the line-of-sight of the galaxies with $E(B-V)>0.2$. This test is consistent with the present calibration.

\begin{figure}
\resizebox{\hsize}{!}{\includegraphics{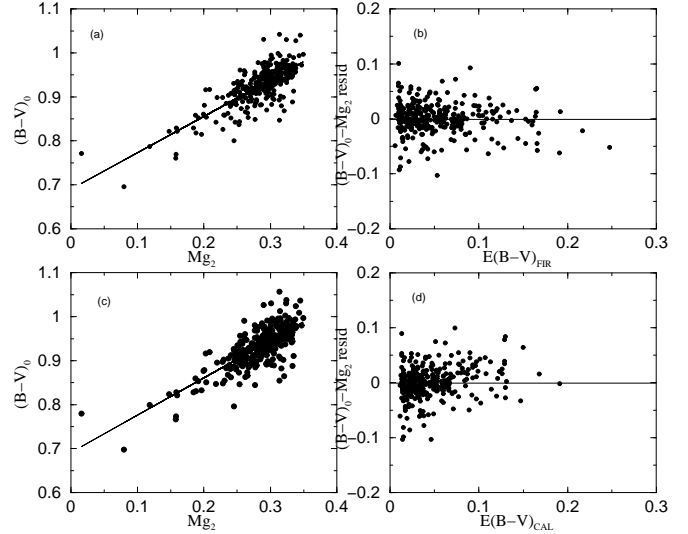}}
\caption{Panels: (a) $(B-V)_0$ {\it versus} $Mg_2$ with $(B-V)_0$ corrected by $E(B-V)_{FIR}$ values ($(B-V)_0=0.823Mg_2+0.689$), (b) residuals from the linear fit in (a) as a function of $E(B-V)_{FIR}$,
(c) $(B-V)_0$ {\it versus} $Mg_2$ with $(B-V)_0$ corrected by $E(B-V)_{CAL}$ values ($(B-V)_0=0.850Mg_2+0.691$), (d) residuals from the linear fit in (c) as a function of $E(B-V)_{CAL}$.}
\label{sample}
\end{figure}

\section{Concluding remarks}

We obtained spectra in the range (3600-6800 \AA) for the nuclear region 
 of 20 early-type galaxies at low Galactic latitudes ($|b|<25^{\circ}$) and total blue magnitude $B_T < 15$ using the Boller \& Chivens spectrograph
 at the CASLEO 2.15-m and ESO 1.52-m telescopes. Two of these galaxies 
 have their radial velocities estimated for the first time.

 We derived  spectroscopically the line-of-sight reddening of these galaxies.
  The reddening in each line-of-sight was
 determined from the comparison of continuum distribution of the galaxy's
 spectrum with that of a reddening-free template with comparable stellar
 population according to the equivalent widths of its absorption features.

We compared the spectroscopic reddening estimates with those derived
 from the 100$\mu$m dust emission ($E(B-V)_{FIR}$) obtained by Schlegel et al. (1998) for the
 lines-of-sight of 54 galaxies, 20 of them from the present work and the remaining 34 from Paper I.
 The comparison reached reddening values up to $E(B-V)$ = 0.65 and indicates that
 for $E(B-V)>0.25$ the dust emission reddening estimates are  higher
 than the spectroscopic reddening values, in agreement with other reddening comparisons 
carried out using different objects. The spectroscopic method proves to be useful
 for reddening  determinations in the line-of-sight of low Galactic latitude galaxies 
 and, therefore, it can be applied using larger telescopes to derive 
 the interstellar reddening in the directions of fainter galaxies
  or of recently 
 catalogued 2MASS galaxies using the infrared domain. The present and future
 reddening estimates using this method should constitute important observational constraints
 to new Galactic extinction models.

We used the present spectroscopic reddening estimates 
(up to $E(B-V) \approx 0.7$) and the SFD98's opacities
 (derived from 100$\mu$m emission and temperature correction maps)
  to recalibrate their dust emission reddening maps. We obtained a calibration factor
 of 0.016 between the spectroscopic $E(B-V)$ values and SFD98's opacities, which is a factor 0.88 lower than that obtained by SFD98. 
 Using the $A_K$ extinction map recently built by 
Dutra et al. (2003) within ten degrees of the Galactic centre, we 
extended the calibration of dust emission reddening maps to low Galactic 
latitudes  down to $|b|>4^{\circ}$ and reddening values of 
$E(B-V)$= 1.6 ($A_V \approx$ 5). According to the new 
calibration, a multiplicative factor of  $\approx$ 0.75 must be 
applied to the COBE/IRAS dust emission reddening maps in those regions.

\begin{acknowledgements} 
We thank the CASLEO staff for hospitality and support during the observing run. 
The authors acknowledge use of the $CCD$ and data acquisition system supported under 
U.S. National Science Foundation grant AST-90-15827 to R.M. Rich. 
We have made use of 
the LEDA database, and the NASA/IPAC Extragalactic 
Database (NED), which is operated by the Jet Propulsion Laboratory, California Institute 
of Technology, under contract with the National Aeronautics and Space Administration. 
We employed  an electronic table from CDS/Simbad (Strasbourg) and  Digitized Sky Survey 
images from the Space Telescope Science Institute (U.S. Government grant NAG W-2166) 
obtained using the extraction tool from CADC (Canada).
 This work was partially supported by the Brazilian institutions FAPESP,
CNPq and FINEP, the Argentine institutions CONICET, Agencia C\'ordoba Ciencia, 
ANPCyT and SECYT (UNC), and the VITAE and Antorchas foundations. CMD acknowledges FAPESP
 for a post-doc fellowship (proc. 2000/11864-6). We acknowledge FAPESP grant 1998/10138-8.
\end{acknowledgements}


\begin{thebibliography}{}
 
\bibitem[]{} Arce, H.G. \& Goodman, A.A. 1999, ApJ, 512, L135
\bibitem[]{} Baldwin, J.A. \& Stone, R.P.S. 1984, MNRAS, 206, 241
\bibitem[]{} Bica, E. 1988, A\&A, 195, 76 
\bibitem[]{} Bica, E. \& Alloin, D. 1986, A\&A, 162, 21 
\bibitem[]{} Bica, E., Alloin, D. \& Schmitt, H. 1994, A\&A, 283, 805
\bibitem[]{} Blakeslee, J.P, Lucey, J.R., Barris, B.J., et al. 2001, MNRAS, 327, 1004
\bibitem[]{} Burstein, D. \& Heiles, C. 1978, ApJ, 225, 40
\bibitem[]{} Burstein, D. \& Heiles, C. 1982, AJ, 87, 1165
\bibitem[]{} Cardelli, J.A., Clayton, G.C. \& Mathis, J.S. 1989, ApJ, 345, 245
\bibitem[]{} Chen, B., Figueras, F., Torra, J., et al. 1999, A\&A, 352, 459
\bibitem[]{} de Vaucouleurs, G., de Vaucouleurs, A. \& Corwin, H.G. 1976 
{\it Second Reference Catalogue of Bright Galaxies}, University of Texas Press (RC2)
\bibitem[]{} Dutra, C.M. \& Bica, E. 2000, A\&A, 359, 347
\bibitem[]{} Dutra, C.M., Bica, E., Clari\'a, J.J., et al. 2001, A\&A, 371, 895 (Paper I)
\bibitem[]{} Dutra, C.M., Santiago, B.X. \& Bica, E. 2002, A\&A, 381, 219
\bibitem[]{} Dutra, C.M., Santiago, B.X., Bica, E. \& Barbuy, E. 2003, MNRAS, 338, 253
\bibitem[]{} Faber, S.M., Wegner, G., Burstein, D., et al. 1989, ApJS, 69, 763.
\bibitem[]{} Ferrari, F., Pastoriza, M.G., Macchetto, F., Caon, N. 1999, A\&AS, 138, 269
\bibitem[]{} Hudson, M.J. 1999, PASP, 111, 57 
\bibitem[]{} Minniti, D., Clari\'a, J.J. \& G\'omez, M. 1989, Ap\&SS, 158, 9
\bibitem[]{} Reach, W.T., Wall, W.F. \& Odegard, N. 1998, ApJ, 507, 507
\bibitem[]{} Sandage, A. 1973, ApJ, 183, 711
\bibitem[]{} Sandage, A. \& Tamman, G. 1981, {\it A Revised Shapley-Ames 
Catalog of Bright Galaxies}, Carnegie Inst. of Washington Publ. 635
\bibitem[]{} Schlegel, D.J., Finkbeiner, D.P. \& Davis, M. 1998, ApJ, 500, 525   
\bibitem[]{} Seaton, M.J. 1979, MNRAS, 187, 73p
\bibitem[]{} van Dokkum, P.G. \& Franx, M. 1995, AJ, 110, 2027

\end{thebibliography}
\end{document}